Urban DNA for cities evolutions
Cities as physical expression of dynamic equilibriums between competitive and cooperative forces

Luca D'Acci

Abstract
Cities are physical manifestations of our competitive and cooperative behaviours. The tension between these two forces generates dynamic equilibriums whose material expressions are cities and their evolutions. In a Darwinian cooperative view, as Darwinism does not involve only competition, the public benefit obtained by cooperation, return in terms of private benefit too. An urban genetic code is proposed, according to which cities emerge connecting nature and urbanity, and as sum of multiuse, independent micro-areas, each one with its centrality, job locations, parks and daily shops-services and amenities. This mechanism, called Isobenefit Urbanism, is not static and pre-designed, but allows infinitely dynamic changes and expansions. Rather than describing *The* ideal city, which doesn't exist outside our own minds, Isobenefit Urbanism describes what a city should avoid to be in order to not become an unideal city. Its six principles are the urban DNA which does not give predetermined forms but indications to follow according to contexts and times. From an environmental angle, Isobenefit cities are resilient, low carbon, adaptive.

1. Competition and Cooperation: an introduction

"Why should you contribute to the public good if free riders reap the benefits of your generosity? Cooperation in a competitive world is a conundrum […] A simple definition of cooperation is that one individual pays a cost for another to receive a benefit. Cost and benefit are measured in terms of reproductive success, where reproduction can be cultural or genetic" (Rand & Nowak 2013, p.413).
Darwin, although often is only associated with a competitive "dog-eat-dog" world and a "rugged individualism", in a book of twelve years later than its *Origin of Species*, wrote: "there can be no doubt that a tribe including many members who were always ready to give aid to each other and to sacrifice themselves for the common good, would be victorious over other tribes; and this would be natural selection" (Darwin 1871).
Models of cultural group selection shown how "in the end, cooperative (pro-social) practices prevail" without involving "necessarily the extinction of groups, but of cultural practices" (Boyd & Richerson 2009).
Many animals live in groups of both kin and non-kin where competitive and cooperative interactions coexist. Examples of altruistic and cooperative behaviour from our closest relatives, and from other social animals as us, abound. "Despite their highly competitive nature, male chimpanzees also cooperate in several contexts to obtain direct and indirect fitness benefits. Coalitionary behaviour, meat sharing, and territorial boundary patrolling provide three example in this regards" (Mitani 2009).
"It is now clear from a variety of animals species that strong, cooperative bonds enhance longevity and offspring's survival. […] chimpanzees do help partners obtain food rewards, especially when the partner is attempting to reach for the food or soliciting help […] chimpanzees do seen motivated to help others and to take into consideration other's outcomes, even when they do not derive direct benefits from doing so" (Cheney 2001, p.10904-5).
With the term cooperation we also include no-inequality. "In humans, inequity aversion, tolerance, and the motivation to engage in joint activities are important catalyst for cooperative behaviour […] Some experiments have suggested that primates rejects food offered by humans if a rival is receiving a better reward [http://www.ted.com/talks/frans_de_waal_do_animals_have_morals from 12:58 to 14:58]. Other studies […] suggest that the food rejections are caused not by perceived inequality but by frustration at seeing, but not obtaining, a preferred food item." (Cheney 2001, p.10905).

Also in economics games, cooperative behaviours are frequent and induce to replace the rationally self-interested *homo economicus* in economics models (Gächter, Renner & Sefton 2008; Fehr & Gächter 2002). Fehr & Shmidt (1999) used a utility curve which include inequality aversion (as a preference expressed for equality among game participants).

Natural/cultural selection acts in two contemporary levels: at the individual level and at the group level.

It is possible that individuals of a group who act in a selfish individualistic way without helping the others, or worse, damaging them, may win in within groups points of view; however in across groups points of view, groups of co-operators may win in a competition with groups of no-co-operators.

Even without assuming competition among groups, when an individual (*x*) helps another (*y*), the latter receives a direct benefit but *x* an indirect benefit too by watching *y* being happier. As a consequence of a fair environment on *y* directly and/or around *y*, *y* may also become more productive, and feel to help others in turn, triggering a positive chain reaction.

Using the words of Adam Smith, "how selfish man may be supposed, there are evidently some principles in his nature, which interest him in the fortune of others, and render their happiness necessary to him, thought he derives nothing from it, except the pleasure of seeing it" (Smith 1759, chapter 1, p.3).

"People cooperate not only for self-interested reasons but also because they are genuinely concerned about the well-being of others […] we came to have these 'moral sentiments' because our ancestors lived in environments, both natural and socially constructed, in which groups of individuals who were predisposed to cooperate and uphold ethical norms tended to survive and expand relative to other groups" (Bowles & Gintis 2011, p.1).

## 2. Competition *versus* Cooperation in Cities

There are different antagonistic forces whose constant tension shapes and changes cities, such as centralization-decentralization, closeness to other and personal space, private profit and public benefit, agglomerative economies and congestion diseconomies.

"Cities exist because individuals are not self-sufficient. If each of us produced everything we consumed and we didn't want much company, there would be ne need to live in cities" (O'Sullivan 2000, p.17).

"We might picture the form of a city as an essential tension between the desire to be as close as possible to everyone else, which is the very idea of a city, and the desire to be as accessible as possible to as much space as possible" (Batty 2013, p.19). The structure of a city, seen as networks and locational patterns, "emerges through countless decisions in the context of physical constraints that limit the feasibility of certain patterns over others. In a sense, economic theory has always assumed that economies develop in this way. Adam Smith's "invisible hand", the life force of capitalism that keeps markets working, is the glue that holds the world together" (Batty 2013, p. 26).

Urban economic theory enables us to understand how competition shapes cities. The economic process of competition is the one which drives city functions, forms, structures and their locations.

Those who are able to pay more gets the best location. If we talk about residences, the one who is able to pay more is the richer; if we talk about activities, it is the one which takes more economical advantage (profit) to specific locations.

The Chicago School's (1920s - 1930s) idea according to which people, like other animals, need to compete to survive is on the basis of the human ecology concept proposed from Robert Park. It analogy to the "ecological struggle for environmental adaptation found in nature", "cities are the outward manifestation of processes of spatial competition and adaptation" (Cooke 1983, p. 133).

This vision was influenced from the Darwinian evolutionary ideas which, translated in urban terms, is underpinned by bid-rent, and suggests that "the most successful urban dwellers would take over the 'best' areas of the city to guarantee their 'survival', while the least successful would end up in less salubrious areas. This process of social competition thus bequeathed a spatial sorting, typically peripheralising the least successful and wealthy urban dwellers" (Hubbard 2006, p. 26).

According to Marx free competition has advantages for the capitalist classes while increases impoverishment of the proletariat which is driven "into even worse and more crowded corners" (Marx 1867, p. 65).

This "free" competition induces only to the rich the capability to choose where to live. On the contrary the poor do not have this capacity, as limited by her/his low monetary budget which excludes her/him to the most expensive areas, which usually are also the most appealing for the ordinary citizen.

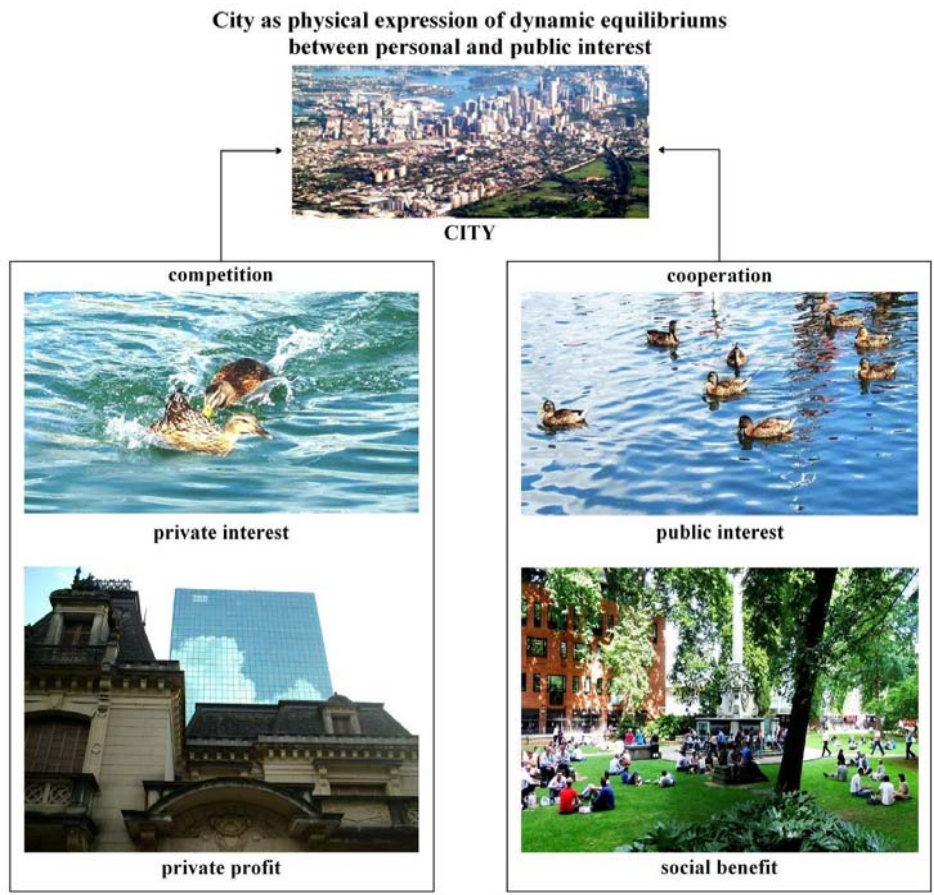

**Fig.1.** Cities as dynamics equilibriums between competitive and cooperative forces.
Source: author's photos

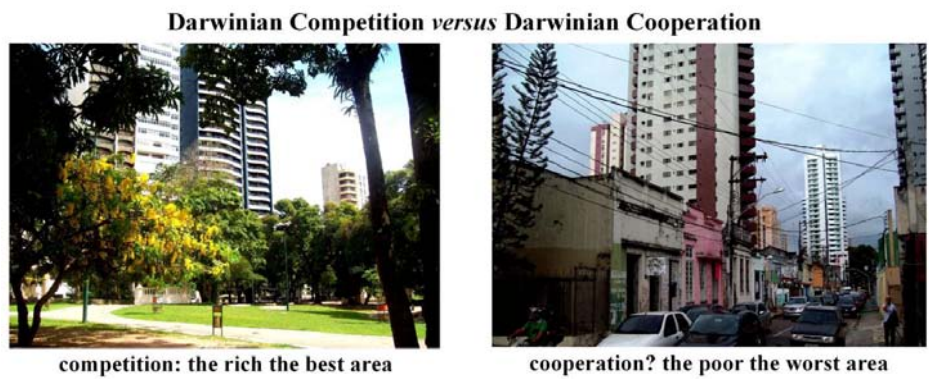

**Fig. 2.** Residential area competition in cities.
Source: author's photos

## 3. An Urban Genotype to balance competition and cooperative forces

Isobenefit Urbanism (D'Acci 2013) proposes an urban planning approach based on the Isobenefit concept.
"Iso" comes from the Greek "ἴσος" meaning "the same", "equal", and is used as a prefix "ἰσο-…".
Therefore Iso-benefit means "equal benefit". The benefit considered here is the one deriving from the advantages which city dwellers receive from the urban morphology, urban structure, transport systems, activities and amenity locations.
The Isobenefit Urbanism approach aims to design cities where each dweller enjoys an equal, or compensative, level of comfort and advantage from the urban quality, services, job locations, urban centers and green.
Its relevance resides in mixing top-down and bottom-up points of view, utopia and complexity, master plans and spontaneity, control and emergence, and in proposing a new idea about the relation between nature and city.
In Isobenefit cities the poor and the rich will enjoy an equal urban quality of the area, services and infrastructures. If you make a city "beautiful" everywhere, you may induce higher housing prices which in turn would push the poor even further away from the city; but the point is that in Isobenefit cities the "even further away" (suburban, sprawl city …) possibility would not exist (land use regulation), therefore only certain parameters would be involved in the mechanism (bottom-up emergence) of the differentiation of prices (i.e. at the building-flat-view level: intrinsic factors) and the "poor" would be able to enjoy a decent and liveable urban area (in terms of services, amenities, green, centralities, job locations, ecosystem services, shops … )
However, even if Isobenefit Urbanism tends to reduce these differences and compensate them (by equalising the relations among distances, amenities and the levels of attractiveness of the latter D'Acci 2015), certainly a location inequality would always be present due to the friction of distance: some points may enjoy the closeness to green areas, centralities and services more than others. Therefore the difference of housing prices will still be in part based on location too (extrinsic factors) and would still allow the poor to find affordable houses inside an Isobenefit city.
In Isobenefit cities each point can reach, with a similar effort, job locations, amenities, centralities, and enjoy an equal level of urban pleasantness. The latter, together with factors such as crime, accessibility, pollutions, services, etc., is also given from Urban Centralities and Fuzzy Urban Quality (D'Acci 2013).
The urban equality intended here refers more to objective features rather than subjective. In fact, while it appears reasonable to say that the desire of having no congestion, no pollution, enjoyment of services and amenities such as pedestrian areas, cleanness, parks, schools and shops may be common for both the poor and the rich, the latter may have different aesthetical and life style preferences.
If we like to write the following Isobenefit Urbanism Principles in a concise way, we could use this expression, which anyway does not add information and the reader could easily jump:

$$\forall k \in \mathrm{M} \quad B_k = \sum_i B_{i,k} = c \geq c^* \quad \wedge F = c \quad \wedge T_{d,k} \leq T^*$$

$$\wedge \exists N \in \vee \notin \mathrm{M} | \forall k \quad d_{N-k} \leq T^*$$

$$\wedge \exists C \in \mathrm{M} | \forall k \quad d_{C-k} \leq T^* \wedge C_1 \neq C_2 \neq \ldots \neq C_n \quad \wedge \exists C | C \gg$$

(1)

The above is read as the following: the city is a matrix (M) of point (k), where $B_{i,k}$ is the benefit in k given from point i: when i is a neutral point, there will be no benefit; when there is an amenity there will be a benefit; when there is a disamenity there will be a negative "benefit".
The overall benefit a point receives from the entire urban points must be constant (c) everywhere for each k. Idem the Fuzzy Urban Quality (F) must be constant in each k of the city (M). The value of this constant (c) must of course be positive and above a certain level of quality ($c^*$).

The time (Td,k) one needs to reach the ordinary daily points from home (work, buying food, recreational time, education, a green area, serene evening walk, agreeable bench…) should not be higher than a reasonable time (T*) which is around 30 minutes walking/10-15 minutes biking (namely within 2-3 km).
Each point must enjoy nature (N), (where we intend real nature, not just a big garden) within T*.
We call C an urban centrality which may be either a particularly pleasant Fuzzy Urban Quality of an area, or a sum of several amenities concentrated in an area, or just an amenity but very attractive. Therefore at least one centrality exists that is reachable from k within T* (in equation 1, dc-k is the "time distance" between C and k).
Each C has a similar overall level of attractiveness, but is not identical in the composition, morphology and location, otherwise we will have a robotic, boring city, without any genius loci throughout its areas, and, in another scale, throughout cities.
Therefore we sketch the Urban DNA of Isobenefit Urbanism:

I. the city morphology and the amenities/disamenities allocation, should guaranty an equal-compensative urban quality for every citizen;

II. each citizen should be able to reach the ordinary daily points within around 30 minutes walking (T*);

III. each citizen should be able to reach a centrality within T*;

IV. each citizen should be able to reach a natural area within T*;

V. buildings citizen should be close to each other in order to create multifunctional, dense and self-sufficient settlements with a sense of urbanity; in the same way natural areas must be dense and large enough to reach a sense of nature (i.e. one square kilometre minimum);

VI. each centrality spontaneously differs from the others, as well as the fuzzy urban quality across areas within the city, and any city elements (architecture, streets shape…);

In addition, small public gardens will be uniformly spread to allow children and seniors to have a green contact by walking within a maximum of 15 minutes rather than the maximum 30 minutes necessary to reach a natural area.
Like the Darwinian rules, Isobenefit Urbanism discourages extensive top-down actions, to favour spontaneous design allowing "chaotic" variety in cities which give cities their specific genius loci.
For this reason Isobenefit Urbanism does not encourage geometrical city forms, ex-novo cities, repetitions and universality. On the contrary, it persuades that forms, locations and functions should emerge from the bottom-up, step by step and according to contexts and preferences throughout history and areas. Its six principles are a kind of urban DNA which does not give any predetermined form and structure; they only offer indications to follow according to contexts. Imagine for example the DNA of a plant which does not directly write any shape, but it just says to it how it should grow with indications such as: build the shape with the goal to take as much light as possible, according to the dominant wind, and towards the strongest possible mechanical resistance in relation to land inclinations and obstacles

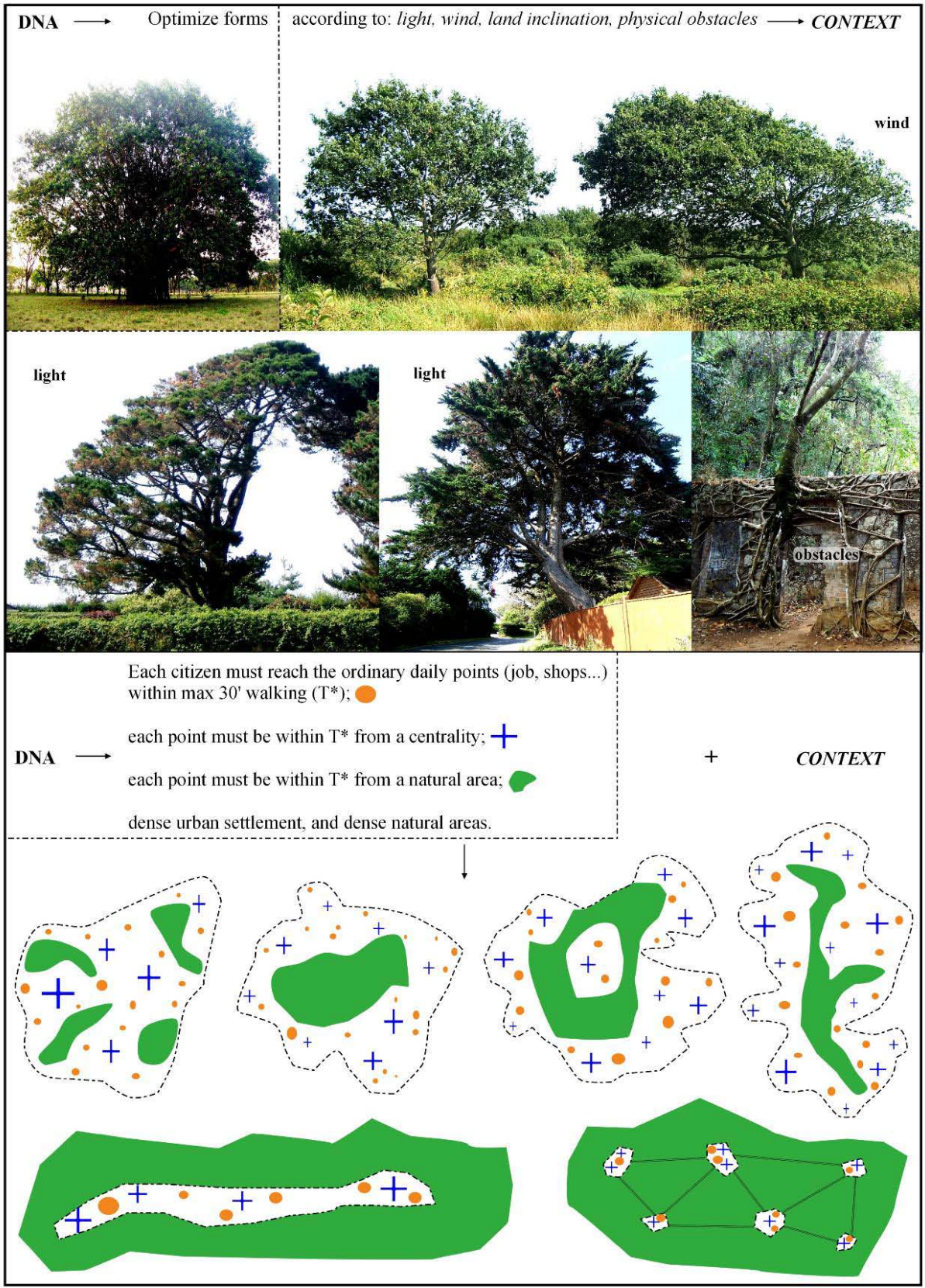

**Fig. 3.** Urban DNA and Context:
Isobenefit Urbanism as *urban genotype* which together with contexts
and time guide spontaneous emergence forms and structures (*urban phenotype*).
Source: author's photos

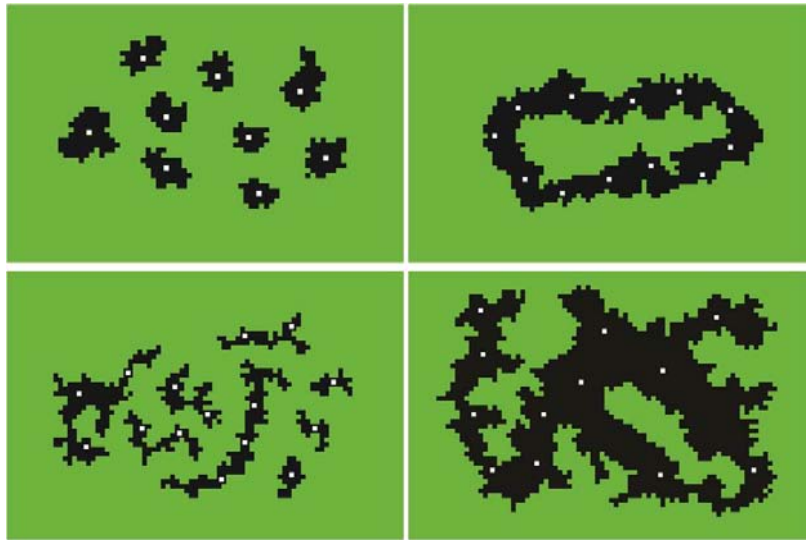

**Fig. 4.** Examples of Isobenefit Urbanism evolutions
(white cells = centralities; green cells= natural areas; black cells= built environment)

We can, at least in part, predict the future (i.e. pollution effects, environment destruction, congested cities, savage urban sprawl, …), therefore the Darwinian natural selection also plays its part on societies or cultural systems which are not able to predict, or give attention to, the long term consequences of their actions.

Talking about cities, when someone (both, top down and bottom up) is certain about objectively negative effects of actions we should drive (top-down together with bottom-up actions) the urban evolution in order to avoid these negative effects already predictable.

The genotype suggested by the Isobenefit Urbanism points, aims to avoid negative and objective effects already predictable, as the following: absence of natural contact and parks, overcrowded traffic, urban heat island effect, GHG and pollutions due to urban morphology and structure, flooding due to excess of continuous cement surfaces.

At another level of reasoning, if we apply the competition Darwinian logic to the residence location competitions, the only one who will take advantage is the rich, as the result is that the poor will be assigned to worse areas within the city, unless they are worse just from the point of view of the rich who therefore will not interfere with their economic value formation as she/he is not interested in living there; namely unless the preferences of the rich and of the poor are different and not overlapping.

However the latter case is not always truth: often the most beautiful area (for aesthetics, quality of design, accessibility, cleanness, modernity, amenities, and/or for other criteria) are attractive for both the rich and the poor, with the difference that the rich can choose them while the poor cannot because the price is fixed from the preferences and willingness-to-pay-for of the rich.

Isobenefit Urbanism suggests to spread quality, accessibility, services and amenities in a uniform, or compensative, way through the city planimetry in order to achieve a comparable attraction and quality across each area within a city helping the poor to live in an equally healthy and attractive urban environment in relation to the rich.

It narrows, in some senses, the urban struggle competition only at the building and flat level (intrinsic factors), and only slightly involving locations too (extrinsic factors).

The Isobenefit Urbanism approach replies in part to the Geddes idea of planning expressed in his "Cities in Evolution" (Geddes 1915), which "was fraught with tension, apparent in the conflict between solving social problems collectively from the top down and the workings of evolution processes which suggest that fitness for purpose emerges from the bottom up" (Marshall & Batty 2009).

Isobenefit Urbanism approaches social problems with a soft top down guide, which is an urban DNA, or the urban genotype but without imposing any final forms and structures; the latter, which we can call the urban phenotype emerge spontaneously following contexts and times.

The Isobenefit Urbanism approach applies the Darwinian urbanism for that which concerns the spontaneous process to build forms and shapes in time and contexts, avoiding universal hyper rational morphologies and geometrical locations of urban objects (parks, residences, shops, services, offices, schools, …) throughout the planimetry of the city, in favour of a more self-organized evolution from where structures and forms emerge from the bottom up rather than from strict and extensive top down planning control and design.

According to post modernism approaches and Geddes's words, Isobenefit Urbanism sees "cities as built environments, inextricable from the societies they housed and the wider natural environment that they were rooted in. In planning terms, this meant that a town [is] not a purely manufactured artefact that could be arbitrarily imposed on a particular location […] but [is] a product of its environment, to be studied as part of that environment, and to be planned in sympathy with it" (Marshall & Batty 2009).

In addition, as we saw, Isobenefit Urbanism is sceptical about allowing a Darwinian residential spatial competition in which, in urban terms, only the rich will take advantage as the ones able to pay most for living in the best areas.

To solve this, it is far distant to suggest a top down allocation of residences and distribution of people, on the contrary, and according to Darwinian urbanism, it still leaves the real estate market and spatial allocation of residences and people free and spontaneous, but it indicates a compensative distribution of centralities, quality and amenities across urban areas. In this way, even in a free context, market and allocations, the poor should be guaranteed for having the capacity to live in a healthy and pleasant area.

As Geddes, rather than emphasising competition, Isobenefit Urbanism emphasises cooperation. "While Geddes accepted Darwin's general theory up to a point, he believed that too much emphasis was given to natural selection and the 'struggle for existence'" (Marshall & Batty 2009).

As other species, we are animals naturally competitive and cooperative and we should consider the community advantage derived from giving a better area to the poor too. The entire community should feel happier and less fragmented (in any terms), and the poor may become also more productive and healthier (physically and mentally) as a consequence of the better underground environment. By an equal spatial distribution of beauties, services and amenities, a lesser feeling of "anger" for inequality treatment may appear too.

Under this cooperative prospective, Isobenefit Urbanism proposes to underline our, still Darwinian in a way, cooperative and altruistic nature too, rather than only the competitive one.

The altruistic behaviour manifestation in cities is providing a good urban environment for the poor too.

Cities are the physical manifestation of our behaviours which are simultaneously competitive and cooperative. The tension between these two forces, competition and cooperation, private interest and public benefit, generate dynamic equilibriums whose material expression are cities.

One aim of Isobenefit Urbanism is then to underline the cooperative behaviour in the sense of giving also to the poor a good urban context where to reside.

Isobenefit Urbanism does not sympathize well with a kind of Social Darwinism involving a "dog-eat-dog" world and a "rugged individualism", but with a cooperative altruistic attitude with the aim to create an equal environment, or at least to reduce extreme differences.

## 4. Conclusion

A way of mixing soft top-down planning with spontaneous bottom-up emergence has been proposed with the idea of conceiving urban DNA codes (urban genotypes) which, together with contexts and times, design cities (urban phenotypes). The urban genotype presented, is, in some senses, still a bottom-up process. In fact it is not a totalitarian pre-designed city plan, but an internal code which allows the emergence of city forms and structures according to our competitive but cooperative behaviours as well, and to time and contexts.

Although this paper presented the Isobenefit urbanism under a social-equalitarian angle, it also (or maybe mostly) shows clear advantages in terms of mitigation (i.e. walkability, public transport), resilience (high ratio green/grey surfaces and relative high water absorption after rains; decentralized spatial location of services, amenities, resources), adaptation (high ratio green/grey surfaces and relative reduction of the urban heat island effect) and quality of life (increased urban quality, contact with nature, pedestrian areas, reduced urban noise and pollution from cars...).

This soft top-down planning (the Isobenefit urbanism genetic code) may still be considered as a bottom-up process because, as Darwin himself underlined, not only competitive, but also cooperative and altruistic behaviours are part of the natural selection, and we have been naturally selected with such instincts from our past natural selection.

We can see Isobenefit cities as asked for, wanted from us (bottom-up), instead of imposed from a top-down actor, and/or as designed step by step along a period of time (bottom-up) instead of all in one go.

If we assume that we all wish for urban quality, walkable cities, reduction of strong inequality, pollution, congestion, increase of multicentrality, multifunctional neighbourhoods, and that we all *need* resilient, adaptive and low carbon environments, and if we like to obtain the human habitat proposed by Isobenefit urbanism, it may appear more correct to define its genetic code as a bottom-up evolution rather than top-down rigid control. It is as if we all self organize our locations, activities, constructions, by auto-checking our emergent urban evolution "from the above" (Isobenefit urbanism code implementation) by asking "someone" (Isobenefit urbanism code) to help us in our self organization process that we want.

## References

Batty, M. (2013). *The new science of cities*. The MIT Press. Cambridge and London.

Body, R. & Richerson, P.J. (2009). Culture and the evolution of human cooperation. *Philosophical Transactions of the Royal Society of London. Series B, Biological Sciences*, 364(1533), 3281-8.

Bowles, S. & Gintis, H. (2011). *A Cooperative Species: Human Reciprocity and Its Evolution*. Princeton University Press, Princeton.

Cheney, D.L. (2011). Extent and limits of cooperation in animals. *PNAS*, June 28, vol. 108 suppl. 2.

Cooke, P. (1983). *Theories of Planning and Spatial Development*. Hutchinson. London.

D'Acci L. (2015). Mathematize urbes by humanizing them. Cities as Isobenefit Landscapes: Psycho-Economical distances and Personal Isobenefit Lines. *Landscape and Urban Planning*. Volume 139, July 2015, Pages 63–81. D'Acci, L. (2013). Simulating Future Societies in Isobenefit Cities. *Futures*. Volume 54, November 2013, Pages 3–18.

Darwin, C. (1859). *On the Origin of Species*. Murray.

Darwin, C. (1871). *The Descent of Man and Selection in Relation To Sex*. Murray.

Fehr, E. & Schmidt, K. (1999). A theory of fairness, competition, and cooperation. *The Quarterly Journal of Economics*, 114(3), 817-68.

Fehr. E. & Gachter, S. (2002). Altruistic punishment in humans. *Nature*, 415(6868), 137-40.

Gachter, S., Renner, E. & Sefton, M. (2008). The long-run benefits of punishment. *Science*, 322 (5907), 1510.

Geddes, P. (1915/1949). *Cities in Evolution: An Introduction to the Town Planning Movement and to the Study of Civics*. Williams & Norgate, London.

Hubbard, P. (2006). *City*. Routledge, London and New York.

Marshall, S. & Batty, M. (2009). From Darwinism to planning – through Geddes and back. *Town & Country Planning*.

Marx, K. (1867) [1887]. *Capital* – Volume One. Progress. Moscow.

O'Sullivan, A. (2000). *Urban Economics*. McGraw-Hill. US.

Rand, D.G. & Novak, M.A. (2013). Human cooperation. *Trends in Cognitive Sciences*, Vol. 17, No. 8.

Smith, A. (1759) *The Theory of Moral Sentiments*.